\documentclass[12pt]{article} 
\usepackage[koi8-r]{inputenc}
\usepackage{cyr}
\usepackage{epsf}
\usepackage{pazha}
\usepackage[round]{natbib}
\usepackage{graphicx}
\usepackage{hyperref}
\tightenlines

\newcommand{\degree}{^\circ}

\usepackage{amsmath} 
\usepackage{xcolor}

\sloppypar

\begin{document}
\baselineskip 21pt

\title{\bf Simulation of images of protoplanetary disks after collision with free-floating planets}

\author{\bf \hspace{-1.3cm}\copyright\, 2024 \ \ 
T. V. Demidova$^{1*}$, V. V. Grigoryev$^1$}

\affil{
{\it Crimean Astrophysical Observatory of the Russian Academy of Sciences, Nauchny, Crimea, Russia}}

\vspace{2mm}
\received{06.09.2024 \\
After revision 29.11.2024; accepted for publication 12.12.2024}

\sloppypar 
\vspace{2mm}
\noindent
Observational manifestations of disturbances in a protoplanetary disk caused by a collision with a massive planet are studied. It is assumed that the planet moves along a parabolic trajectory that intersects the disk plane near the star. Gas-dynamic simulation is performed using the finite volume method on a long time scale. On its basis, images of the disk observed from the pole and edge-on are constructed in the infrared and submillimeter ranges. A wide range of planet orbit parameters is considered. The approach of the planet was considered both prograde and retrograde with the respect to the disk rotation. Calculations have shown that in the images of the disk seen pole-on, two spiral arms can be observed in case of the prograde fall, and one with retrograde case. In the case of observations of a disk whose plane is inclined at a small angle to the line of sight, distortions of the disk plane can be noticeable. In addition, a gas tail is extended from the disk in the direction of the planet's motion, which can also be identified in observations. 


\vfill
\noindent\rule{8cm}{1pt}\\
{$^*$ email: proxima1@list.ru}

\clearpage

\section*{INTRODUCTION}
\noindent

Free-floating planets are planetary-mass objects that are not gravitationally bound to any star. Such objects have been identified in star cluster surveys~\citep{2000Sci...290..103Z, 2006MNRAS.373L..60L,2010ApJ...708..770Q,2012A&A...548A..26D,2012ApJ...756...24S,2012ApJ...754...30P} and by gravitational lensing in the Galactic field~\citep{2011Natur.473..349S, 2012ARA&A..50..411G,2017Natur.548..183M}. An interstellar planetary-mass object has also been discovered in the $\beta$ Pic group of moving stars~\citep{2013ApJ...777L..20L}. Recently, isolated objects comparable in mass to Jupiter, named JuMBOs, have been discovered using the JWST telescope~\citep{2023arXiv231001231P}. 

According to current estimates, the number of free-floating planets with a mass of several Jovian masses in our Galaxy should exceed the number of Main Sequence stars by at least two times~\citep{2011Natur.473..349S,2013ApJ...778L..42P}. \citet{2018MNRAS.473.1589G} showed that $\sim 1 \%$ of low-mass stars ($<2M_\odot$) capture a free-floating planet during their lifetime. Such planets can be located on distant orbits from the star ($>100$~AU), have a noticeable eccentricity and a significant orbital inclination relative to the equatorial plane of the star, and also move retrograde relative to other planets in the system.

Various mechanisms have been proposed for the formation of free-floating planets. First, free-floating objects of planetary mass may form during the gravitational collapse of a protostellar cloud~\citep{2001ApJ...551L.167B,2004ApJ...617..559P}. The number of such objects in populations found in several star clusters agrees well with the extrapolation of the initial mass function of stars to the planetary mass region~\citep{2009A&A...508..823B,2010ApJ...719..550M,2012ApJ...756...24S, 2015ApJ...810..159M}. Such objects add to the family of large-mass free-floating planets.

Secondly, at the stage of formation of the planetary system, massive Jupiter-like planets can be arranged in such a way that the system becomes dynamically unstable. In this case, one of the planets moves to a highly eccentric orbit close to the star (which can subsequently be rounded by tidal interaction), and the second planet is ejected from the system into interstellar space~\citep{1996Sci...274..954R}. In~\citet{2008Icar..193..475M} a system with two giant planets was considered, which, migrating in the protoplanetary disk, can disturb each other's orbits, with a subsequent increase in the eccentricities of their orbits. Such systems are unstable, and often the result of the gravitational interaction of the giant planets is the ejection of the planet from the system or its fall onto the star.

Third, clumps of matter formed due to gravitational instability in a massive protoplanetary disk can be ejected from it into interstellar space after approaching another star~\citep{2012ApJ...750...30B,2020A&A...635A.196V}. The ongoing collapse of such clumps can lead to the emergence of free giant planets.

Another source of free-floating planets may be ejections during the approach of the formed system to another single or binary star.
In~\citet{2011MNRAS.411..859M} it is shown that during a close flyby ($r_{\rm min}<100$~AU) one or more planets can be instantly ejected from the system, such planets can be captured by the incoming star or remain free-floating. In addition, the approach to an external star can destabilize the system, which, in turn, will lead to a delayed ejection of the planet. The eccentricities of the planets can increase during multiple approaches at distances of $100$~AU$<r_{\rm min}<1000$~AU. Massive numerical calculations~\citep{2015MNRAS.448..344L}  showed that the interaction cross-section sufficient for the ejection of a planet increases with its semimajor axis and eccentricity. It decreases with the growth of the mass of the parent star and the velocity dispersion of neighboring stars. At the same time, the interaction cross-section sufficient for ejection is smaller in the case of approaching a single star than a binary star.

Approaches of stars at distances sufficient for the ejection of planets from their planetary systems are most likely in star clusters~\citep{2009ApJ...697..458S,2019MNRAS.489.2280F}. Under the gravitational influence of neighboring stars (single and binary), the eccentricities of planets can increase significantly~\citep{1998ApJ...508L.171L,2004AJ....128..869Z}, some outer planets can leave the system~\citep{2012ApJ...754...57B,2013MNRAS.433..867H}, and also leave the star cluster itself. The number of such planets increases with time~\citep{2015MNRAS.453.2759Z}. Changes in the orbital elements of planets contribute to the destabilization of their orbits relative to each other, and this increases the probability of the ejection of a planet from a planetary system~\citep{2017MNRAS.470.4337C}.

Planets with orbital elements and Jovian mass can maintain orbital stability and protect the inner (relative to its orbit) planets of the system from external interactions, in the case of a small number of neighboring stars~\citep{2019MNRAS.489.2280F}. However, with a high concentration of external stars, the orbit of a Jupiter-type planet can be noticeably destabilized. The location of the planets in resonances can prevent the destruction of the system and the ejection of giant planets~\citep{2020MNRAS.497.1807S}. The consequences of the passage of a body with the mass of a planet or a brown dwarf through the Solar System were considered in~\citet{2024MNRAS.528.6411M}. According to this paper, the Solar System has not experienced close encounters with free-floating planets of Jupiter mass or more during its entire existence, otherwise it would have been destabilized. Thus, the probability of such interaction for an isolated planetary system is extremely small. However, it is impossible to exclude such interaction at the stage of the protoplanetary disk with a large number of neighboring stars in the cluster.

In~\citet{2020AJ....160..232Z} shows that at late stages of evolution, Sun-type stars can lose most of their mass, which will lead to the removal of the orbits of giant planets. In this case, a Solar-type system will be less resistant to disturbances during the passage of an external star, and the ejection of all giant planets except one is possible. Probably, catastrophic events occurring at the last stages of the evolution of stars of different masses should also contribute to the ejection of planets into interstellar space.

Considering the many mechanisms by which free-floating planets can form and the multiple observations of such objects, it can be assumed~\citep[as noted in][]{2011Natur.473..349S} that the number of such objects is at least several times greater than the number of stars in the Galaxy. Therefore, the intrusion of such an object into the immediate environment of a star is not an exceptionally rare event during the lifetime of a protoplanetary disk in the presence of a dense stellar environment, for example, in star clusters.

In this paper we consider the consequences of a massive planet's passage through the protoplanetary disk material. The possibility of observing the inhomogeneities of the disk material that arise in such a case is demonstrated and investigated.

\section*{MODEL AND METHOD}
\subsection*{Basic equations}
The calculations were performed using the PLUTO package for numerical solution of gas dynamics problems~\citep{2012A&A...545A.152M}. A system of non-stationary gas-dynamic equations was considered that describe the evolution of a gas disk surrounding a young solar-type star in a spherical coordinate system
$(R, \theta, \varphi): 144 \times 60 \times 144$ cells in area $[0.2; 107.2]$~AU $\times  [15; 165]\degree$ $\times [0; 360)\degree$.

Let's write the continuity equation:
\begin{equation} \label{eq:masscons}
	\displaystyle \frac{\partial \rho}{\partial t} + \nabla \cdot \left( \rho \vec{v} \right)  =  0,
\end{equation}
where the gas density is denoted by $\rho$, and the velocity is $\vec{v}$.
Let us write the equation of gas motion:
\begin{equation} \label{eq:navierstokes}	
	\displaystyle\frac{\partial (\rho \vec{v})}{\partial t} + \nabla \cdot \left( \rho \vec{v} \cdot \vec{v} -  p \hat{I} \right)^T =  - \rho \nabla \Phi + \nabla \cdot {\sf \Pi}(\nu).
\end{equation}
Here: $p$~is gas pressure, $\hat{I}$~is identity matrix, $\Phi = \Phi_\ast+\Phi_p$, where $\Phi_\ast=-GM_\ast/R$~is gravitational potential created by the star ($G$~is gravitational constant, $M_\ast = 1 M_{\odot}$~is mass of the star, $R$~is distance to the star, $M_{\odot}$~is mass of the Sun, calculations are made in the coordinate system relative to the star), and $\Phi_p$~is smoothed gravitational potential of a planet with mass $m_p$, which is located at point $\vec{R}_p$~\citep{2006A&A...445..747K}:
\begin{equation}
\Phi_p = 
\begin{cases}
	-\frac{Gm_p}{d}, &  d > R_{sm} \\
	-\frac{Gm_p}{d} \left( \left(\frac{d}{R_{sm}}\right)^4 - 2\left(\frac{d}{R_{sm}}\right)^3 + 2 \left(\frac{d}{R_{sm}}\right) \right), & d < R_{sm}
\end{cases}
\end{equation}
in this case the parameter $R_{sm} = 0.03R$, $d = \sqrt{(\vec{R}-\vec{R}_p)^2}$. ${{\sf \Pi}}(\nu)$~is viscous stress tensor:
\begin{equation}
	{\sf \Pi}(\nu) = \nu_1 \left[ \nabla \vec{v} + (\nabla \vec{v})^T\right] + \left(\nu_2 - \frac{2}{3}\nu_1 \right)(\nabla \cdot \vec{v})\hat{I},
\end{equation}
$\nu_1$~is kinematic viscosity coefficient, $\nu_2$~is second viscosity ($\nu_2=0$).

Let us write the energy equation~\citep[similar to][]{2012A&A...545A.152M}:
\begin{equation} \label{eq:energycons}
	\displaystyle\frac{\partial \varepsilon_t}{\partial t} + \nabla \cdot \left[ \left(\varepsilon_t + p\right)\vec{v}   \right]  = \nabla \cdot (\vec{v} \cdot {\sf \Pi}(\nu)) + \nabla \cdot \vec{F}_c - \rho\vec{v}\nabla\Phi.
\end{equation}
Here the total energy density $\varepsilon_t = \rho \epsilon + \rho \vec{v}^2/2$ includes the specific internal energy of the gas $\epsilon$. The heat flux is determined through the thermal conductivity coefficient $\kappa$ and the gas temperature gradient $T$: $\vec{F}_c = \kappa \cdot \nabla T$. The coefficients for viscosity and thermal conductivity are specified taking into account the turbulent and laminar terms by analogy with~\citep{2024ARep...68..949G}. The use of non-zero transport coefficients is necessary to smooth out the narrow transition layer between the rarefied hot corona of the disk and the denser cold matter of the disk. In addition, when the planet moves in a gaseous medium, the appearance of a bow wave is expected, which can potentially lead to unnecessary numerical discontinuities that cannot be resolved with the computational grid used. The presence of thermal conductivity and viscosity also makes it possible to smooth out this issue.   

The equation of state of an ideal gas closes the system of equations (\ref{eq:masscons})--(\ref{eq:energycons}):
$p = \rho \epsilon (\gamma - 1)$, $\gamma = 1.05$, so that the temperature remains nearly constant. This approximation
simulates the situation where molecular and atomic gases remain nearly isothermal. It is assumed that a given temperature profile is maintained in the disk volume throughout the simulation.

It should be noted that this paper did not consider the influence of processes associated with radiation transfer (heating by UV radiation from hot regions of the star, cooling of dust) and the interaction of dust and gas. Self-gravity of the disk was also not taken into account.

\subsection*{Initial and boundary conditions, units of measurement}
Gas-dynamic calculations are performed in the system of units listed in Table 1.

\begin{table}[t]
\vspace{6mm}
\centering
{{\bf Table 1.} Normalization units} \label{tab:dimensionalization}
\vspace{5mm}\begin{tabular}{|l|c|c|c|l|}
			\hline\hline
			Parameter           & Designation & Value               & in CGS units  & Comment     \\[2pt] \hline
			Unit of mass     & $M_0$       & $1.98 \times 10^{33}$  & g          & $M_{\odot}$     \\
			Unit of length     & $L_0$       & $1.496 \times 10^{13}$ & cm         & $1$~AU        \\
			Unit of time   & $t_0$       & $3.16 \times 10^7$     & s          & $1$~year         \\
			Unit of density & $\rho_0$    & $5.94 \times 10^{-7}$  & g cm$^{-3}$ & $M_0/L_0^3$     \\
			Unit of velocity  & $v_0$       & $4.74 \times 10^{5}$   & cm s$^{-1}$       & $2 \pi L_0/t_0$ \\
			\hline
		\end{tabular}
\end{table}

Following~\citet{2013MNRAS.435.2610N}, the initial density distribution in the disk was given by:
\begin{equation}\label{eq:rho}
	\rho(r, z, 0) = \rho_0\left(\frac{r}{r_{\rm in}}\right)^p\exp\left(\frac{GM}{c_s^2(r)}\left[\frac{1}{\sqrt{r^2+z^2}}-\frac{1}{r} \right]\right).
\end{equation}
Here $r = R\sin \theta$~is cylindrical radius, $z$~is height above the equator, $c_s(r)$~is speed of sound at this radius, $r_{\rm in}$~is inner radius of the disk.

The speed of sound in the disk was determined as follows
\begin{equation}\label{eq:cs}
c_s^2(r)=c_0^2\left(\frac{r}{r_{\rm in}}\right)^q,
\end{equation}
where $c_0$~is the speed of sound at a distance $r_{\rm in}$. The speed of sound together with the angular Keplerian velocity
$\Omega_K(r)=\sqrt{GM/r^3}$ determines the disk half-thickness $H(r)=c_s(r)/\Omega_K(r)$. From equation~(\ref{eq:cs}) it follows that
\begin{equation}\label{eq:H}
H(r)=H_0\Big(\frac{r}{r_{\rm in}}\Big)^{(q+3)/2},
\end{equation}
where $H_0=c_0/\Omega_K(r_{\rm in})$.

In this paper, the values $p=-2.25$ and $q=-0.5$ are adopted. In this case, the surface density distribution is determined by the formula $\Sigma\approx\Sigma_0\left(\frac{r}{r_{in}}\right)^{-1}$. The value of $\Sigma_0$ is determined by the total mass of the disk, which is equal to $M_d=0.01M_\odot$~\citep{2011ARA&A..49...67W}. In this case, $\rho_0=\frac{\Sigma_0}{\sqrt{2\pi}H_0}$. The value of $c_0$ was chosen so that the temperature at the inner boundary of the disk ($r_{\rm in}=0.2$) was equal to $T_0=\frac{c_0^2\mu m_H}{\gamma k_B}=1000$~K, in which case $H_0\approx 0.006$~AU. The molar mass of the gas in the disk is $\mu = 2.35$~\citep{1994A&A...286..149D}, $m_H$~is the mass of the hydrogen atom, and $k_B$~is the Boltzmann constant.

The density in the cell is limited from below by the value $10^{-10} \rho_0$, and the entire region filled by such a rarefied medium is considered as a corona with the temperature of $10^4$~K and a molar weight of $\mu=0.71$.

The initial angular velocity of the substance is given by the formula:
\begin{equation}\label{eq:Om}
\Omega(r,z)=\Omega_K(r)\left[(p+q)\left(\frac{H}{r}\right)^2+(1+q)-\frac{qr}{\sqrt{r^2+z^2}}\right]^{1/2}.
\end{equation}
Thus, the coordinates of the velocity of each cell of the disk are set equal to $\vec{v} = (v_R, v_{\theta}, v_{\varphi}) = (0, 0, \Omega r)$. In the case of calculations with a retrograde fall of the planet, the initial azimuthal velocity is replaced by a negative one: $v_{\varphi} = - \Omega r$. Thus, with a retrograde fall of the planet, the disk rotates clockwise in the $xy$ plane, and with a prograde fall, it rotates counterclockwise.

The left boundary condition on $R$ is set from considerations of equality of thermal turbulent fluxes, continuity of viscous turbulent flow and conservation of angular momentum. The values in the boundary cells are designated by the subscript $b$, without the subscript~are in the calculated boundary cells:
\begin{equation}
	\kappa_{turb~b} \frac{\partial T_b}{\partial R} = -\kappa_{turb} \frac{\partial T}{\partial R}; \qquad \nu_{turb~b} \frac{\partial \rho_b \vec{v}_b}{\partial R} = \nu_{turb} \frac{\partial\rho \vec{v}}{\partial R}; 
	\qquad 
	\rho_b v_{\varphi~b} R_b = \rho v_{\varphi} R. 
\end{equation}
in this case, the component $v_{Rb}$ is always directed towards the star or is zero.

The right boundary condition on $R$ is free, i.e. all quantities used in the calculations have a gradient on $R$ equal to zero. Free boundary conditions on $\theta$ are also set on both sides and periodic boundary conditions on $\varphi$.

At the initial moment of time, the planet with the mass $m_p = 10 M_{Jup}$ ($M_{Jup}$~is Jupiter's mass) is located at a distance of $R_{p0}=150$~AU. Calculations showed that the disk matter is disturbed by the planet when it flies directly above the disk at a distance of $\leq 100$~AU from the star. Doubling the initial distance to the star did not have a significant effect on the behavior and lifetime of the observed structures. The initial velocity of the planet was taken to be equal to the free fall speed at this distance $V=\sqrt{\frac{2GM_\ast}{R_{p0}}}$. The longitude of the ascending angle of the planet's orbit in all calculations was set at $\Omega=0\degree$ (the intersection of the disk plane occurs on the $x$ axis), the pericentric distance ($q$), the argument of the pericenter ($\omega$), and the inclination ($i$) of the planet's orbit {to the initial plane of the disk} were varied.

\subsection*{Image generation}
For three-dimensional calculations of radiative transfer with a known gas density distribution in the region under consideration,
the RADMC-3D code was used \footnote {https://www.ita.uni-heidelberg.de/dullemond/software/radmc-3d/}~\citep{2012ascl.soft02015D}, in which the solution of the radiative transfer equation is performed by the Monte Carlo method.
In this paper, we assume that the total mass of fine dust ($0.1$~$\mu$m) is equal to $\sim 2\times 10^{-5} M_\odot$.
The amount of fine dust decreases during the evolution of the protoplanetary disk~\citep[see, e.g.,][]{2014prpl.conf..339T} 
due to the enlargement of dust grains. We assume that the maximum dust size less than the Rayleigh limit ($\lambda \approx 2\pi a_{\rm max}$, where $\lambda$~is wavelength, $a_{\rm max}$~is radius of a spherical dust grain), at which a jump in dust opacity occurs, is not reached. $10^8$ photons were used for calculating direct and diffuse radiation. Dust opacity for magnesium-iron silicates (Dorschner et al., 1995) was calculated using the Mie theory~\citep{1908AnP...330..377M} using the code included in the RADMC-3D package~\citep{1998asls.book.....B}. 

The calculated radiation fluxes were used to simulate images that could potentially be obtained from observations with the ALMA radio interferometer. The simulation was performed using the CASA 6.5 simulator~\citep{2012ASPC..461..849P}. The calculations were performed at a wavelength of 740 $\mu$m (Band 8) for the position $\alpha= 05^{\rm h} 35^{\rm m} 58^{\rm s}.5; \delta = +24^\circ 44' 54''.1$ (CQ Tau) at a distance of 140 pc, the bandwidth for the continuum observer is $6$ GHz, the exposure time is $2^{\rm h}$. The antenna configuration was Cycle 5 (5.7 of the available CASA configurations), with a beam size of $\approx 0''.15$. Thermal noise was added using the tsys-atm option of the CASA package, with a precipitable water vapor of $PWV = 0.6$. 

\section*{RESULTS}
\noindent
In all models, the planet moves along a parabolic orbit from the lower part of the half-space ($z<0$), crosses the disk plane ($z=0$) twice and goes under the disk again. The planet's motion gravitationally disturbs the disk matter. Therefore, as shown by gas-dynamic calculations, the longer the planet is in close proximity to the plane of the protoplanetary disk, the more it is distorted, and the more noticeable are the large-scale asymmetries in its images. Long-term gas-dynamic calculations of the model with parameters $q=5$~AU, $i = 10\degree$ and $\omega= 90\degree$ for the prograde and retrograde flyby of the planet were selected, on the basis of which disk images were constructed from the pole and from the edge in the near infrared (IR) region of the spectrum (3~$\mu$m). The orbit of such a planet passes the pericenter point at the time $t\approx 144$~year from the start of the calculations.

\begin{figure}[!ht]
\centering\includegraphics[width=0.9\textwidth]{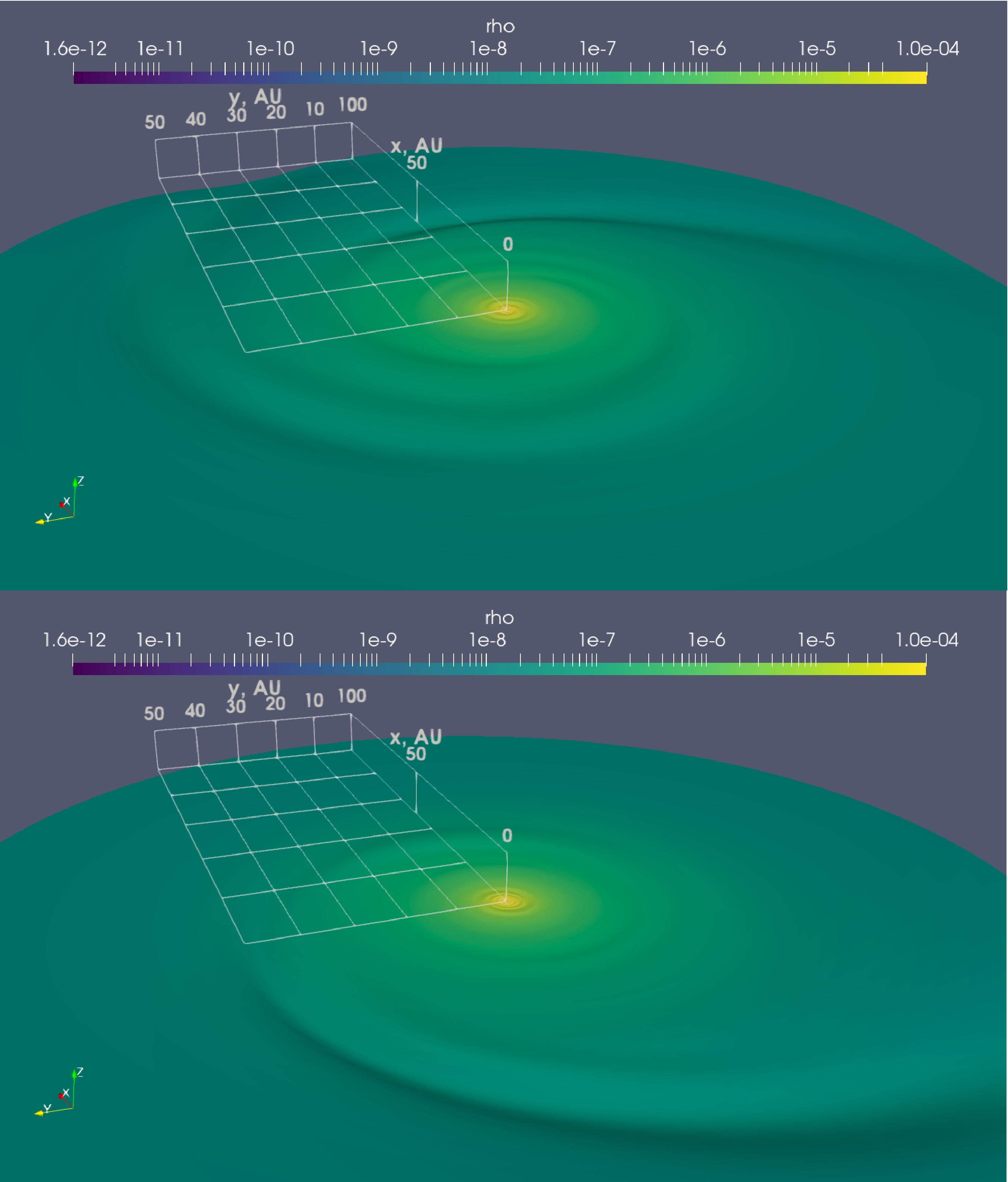}
\caption{\rm Surface of maximum density at the time $290$~years relative to the start of calculations in the model $q=5$~AU,
$\omega=90\degree$, $i=10\degree$. The case of a prograde flyby is shown at the top, and a retrograde one at the bottom. The color shows the density in units of $\rho_0=6\times 10^{-7}$ g cm$^{-3}$. The coordinate parallelepiped is shown for scale.}\label{fig:density}
\end{figure}

Calculations have shown that the plane of the disk warps, and in the direction of the planet's motion a hump of matter is formed, which shifts along the disk due to the Keplerian rotation. Then, after the second intersection of the plane of the disk, a second spiral hump should form. And it does form in the case of a prograde fall of the planet (Fig.~\ref{fig:density}, top). However, in the case of a retrograde fall, the second hump merges with the first due to the rotation of the disk (Fig.~\ref{fig:density}, bottom). The density of matter in the spiral humps exceeds the density of the surrounding medium of the disk. For example, at a distance of $40$~AU at a time of 200 years, the density in the hump is 1.7 times higher than the density at the same distance in the plane of the disk, whereas after 290 years is 1.25 times. The temperature of the dust on the surface of the hump closest to the star is several degrees ($2 - 5$~K) higher than that of the surrounding medium.

\begin{figure}[!ht]
\centering\includegraphics[width=0.9\textwidth]{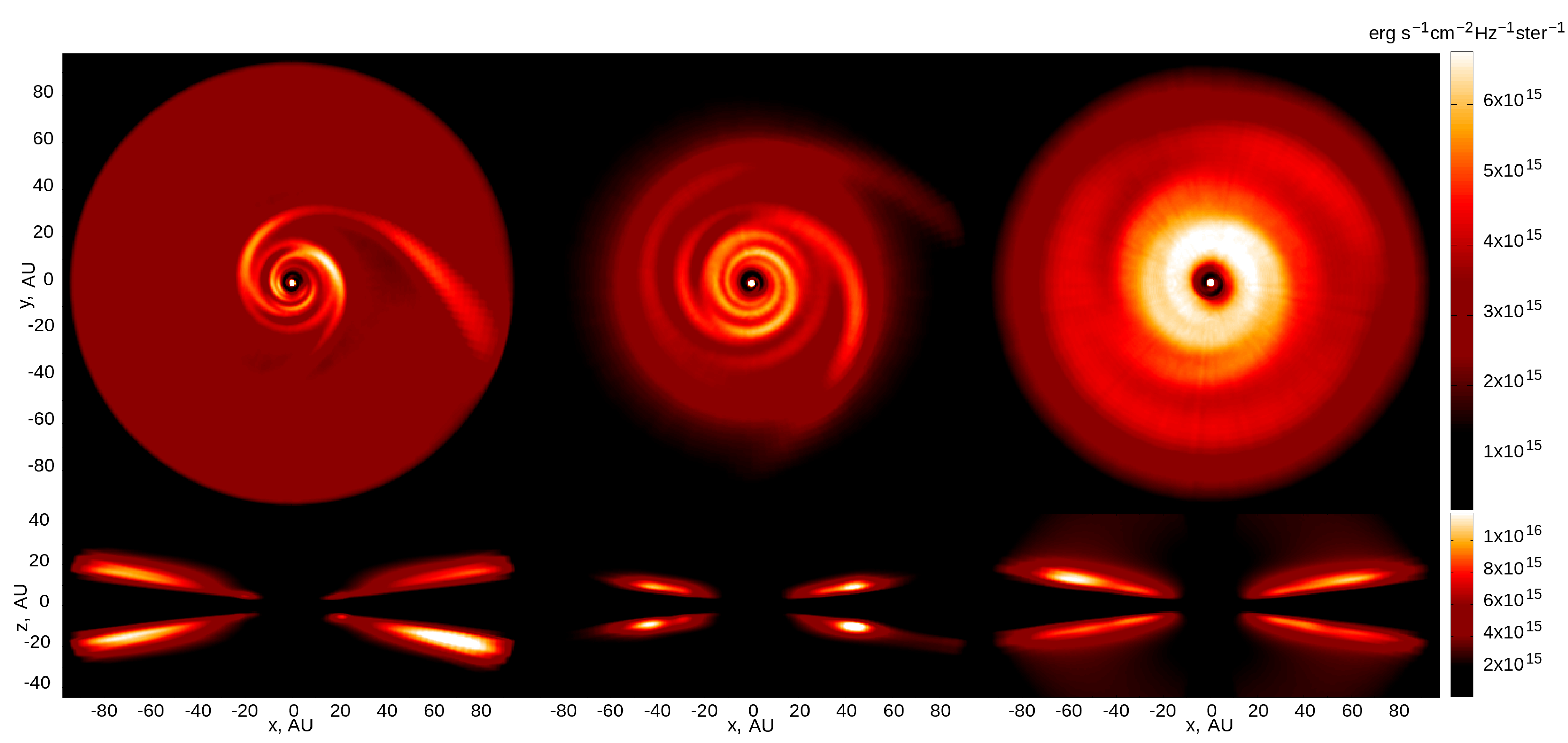}
\caption{\rm The color shows the product of the radiation flux at a wavelength of 3~$\mu$m and the square of the distance for the case of a prograde fall of the planet onto the disk. Model parameters: $q=5$~AU, $\omega=90\degree$, $i=10\degree$. The time moment $t=200$~years from the beginning of the calculations is shown on the left, $t=290$~years is shown in the center, and $t=500$~years is shown on the right. The upper graphs correspond to the direction of the line of sight against the $z$ axis, and the lower ones along the $y$ axis.}\label{fig:timeP}
\end{figure}

Therefore, when the planet approaches the pericenter point, a single-arm spiral is visible in the IR image of the disk, which is a spiral hump protruding under the disk. Then, during a prograde flyby of the planet, a double-arm spiral is visualized in the image of the protoplanetary disk, seen pole-on (Fig.~\ref{fig:timeP}). Inhomogeneities in the disk atmosphere are visible in the edge-on image. In the case of a retrograde fall of the planet, only a single-arm spiral is observed in the outer part of the disk, which can also be seen in edge-on observations, especially in the early stages of the planet's interaction with the disk (before passing the pericenter and shortly after).

In the inner part of the disk in both cases bright two-armed spirals are visible. Apparently, their origin is connected with the fact that the planet crosses the plane of the disk twice. The direction of twisting of the spirals is opposite for retrograde and for prograde flybys of the planet. The decrease in the visible size of the disk at the moment of $290$ years is connected with the bending of the disk due to the reaction to the flyby of the planet.

\begin{figure}[!ht]
\centering\includegraphics[width=0.6\textwidth]{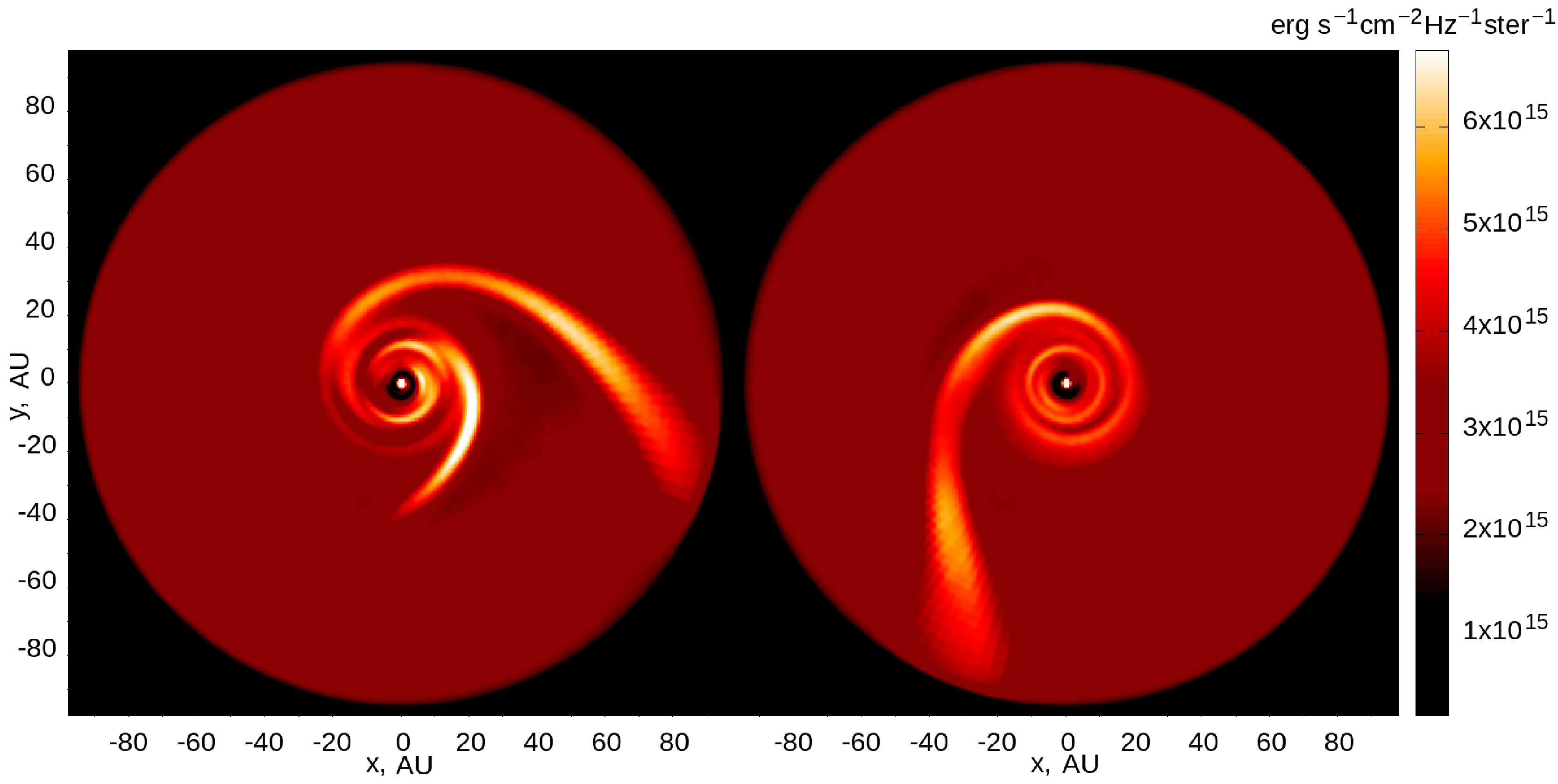}
\caption{\rm The same as in Fig.~\ref{fig:timeP} for the case when the line of sight is directed along the $z$ axis at time $t=200$~years.
On the left is a prograde flight, and on the right is a retrograde one.}\label{fig:i180}
\end{figure}

\begin{figure}[!ht]
\centering\includegraphics[width=0.9\textwidth]{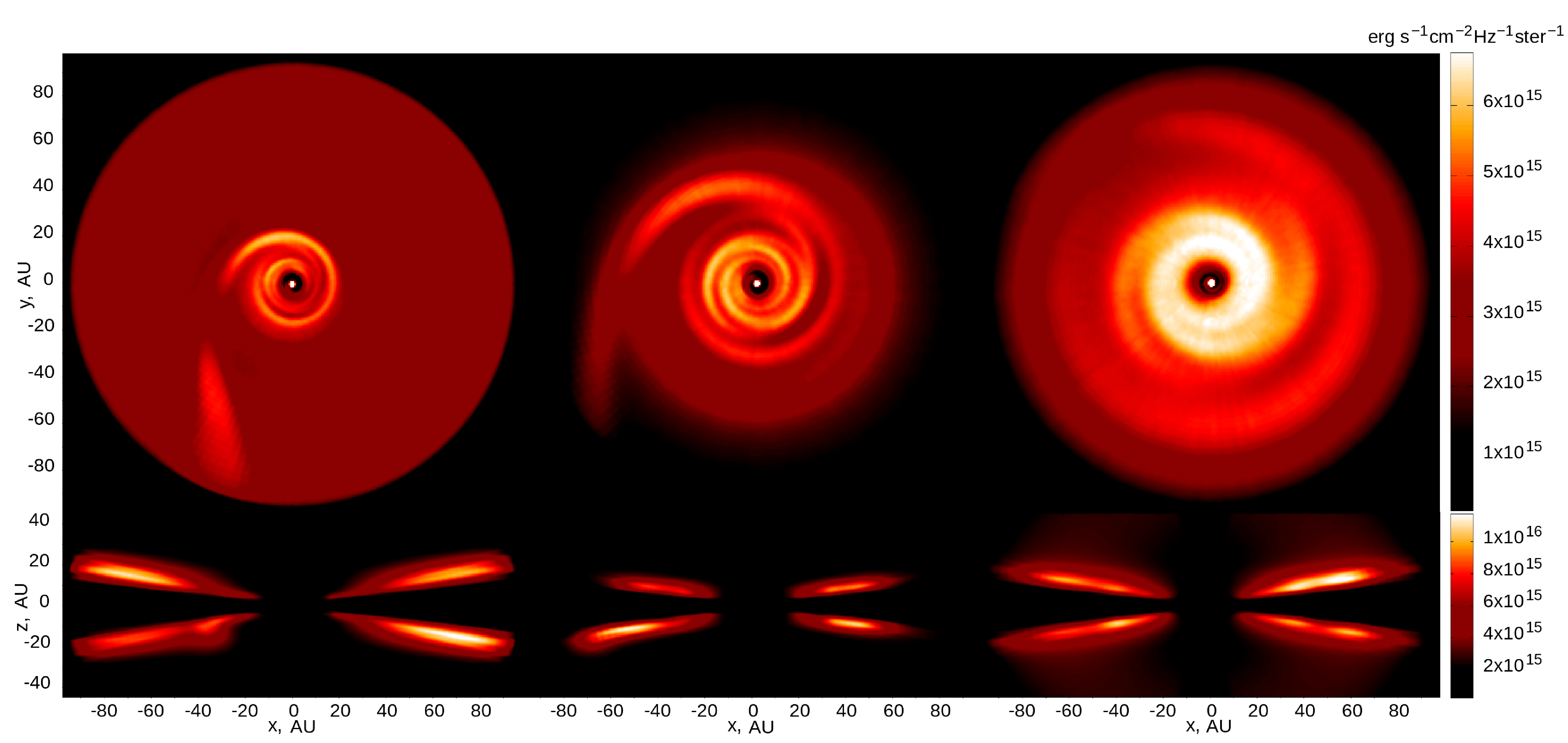}
\caption{\rm The same as in Fig.~\ref{fig:timeP} for the case of a retrograde fall of the planet onto the disk.}\label{fig:timeR}
\end{figure}

Fig.~\ref{fig:i180} shows images of the disk observed from below in the direction of the $z$ axis at time $t=200$~years. It is evident that the structures are brighter compared to observations from above (Fig.~\ref{fig:timeP}, \ref{fig:timeR}). This is due to the fact that in this case the humps of matter are located under the plane of the disk (closer to the observer), since the planet approaches the disk from below ($z < 0$). Differences are noticeable when the planet is located near the pericenter and before it in the IR range. Then the matter is vigorously mixed due to the rotation of the disk, and already at time $t = 290$~years the view from below and from above are practically indistinguishable from each other.

Over time, the disturbance covers the entire disk, and by the time $t \approx 290$~years, when the planet moves away to a distance equal to the initial one, a large-scale spiral structure is visible in the images from the pole, and an asymmetry relative to the plane of the disk is noticeable in the edge-on observations. After $500$~years from the start of the calculations, the spiral structure is still noticeable in the disk, but the picture looks blurry (Fig.~\ref{fig:timeP},~\ref{fig:timeR}). 

\begin{figure}[!ht]
\centering\includegraphics[width=0.6\textwidth]{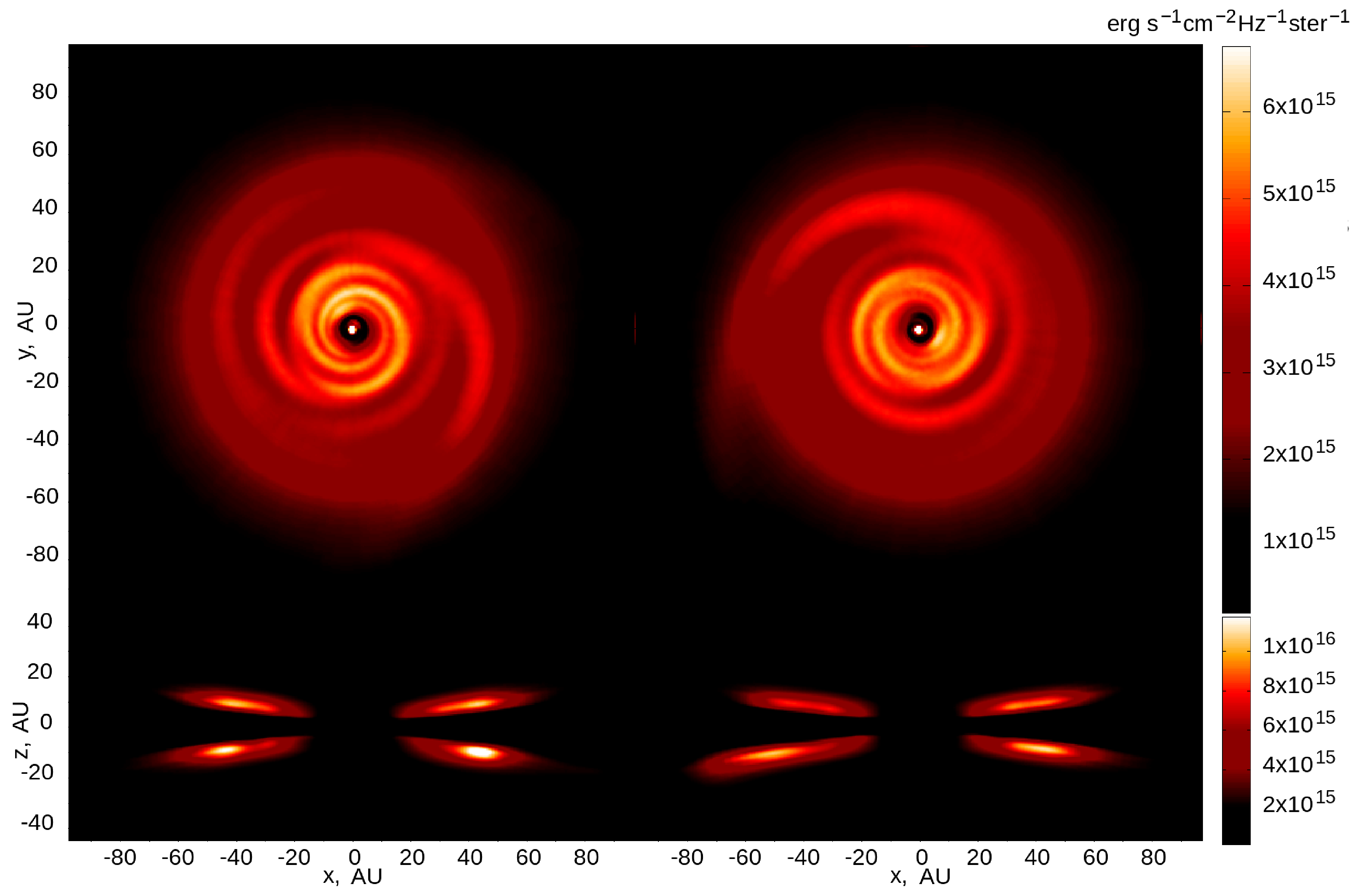}
\caption{\rm Same as in Fig.~\ref{fig:timeP} for the inclination of the planet's orbit $i=20\degree$ ($q=5$~AU, $\omega=90\degree$) at
time $t=290$~years. On the left is the prograde flyby, and on the right is the retrograde one.}\label{fig:incl}
\end{figure}

As the inclination of the planet's orbit increases, the time it spends in the immediate vicinity of the disk plane decreases. Fig.~\ref{fig:incl} shows disk images for the model with parameters $q=5$~AU, $i = 20\degree$ and $\omega = 90\degree$ for the prograde and retrograde cases. Spiral structures are still visible, but they are less bright, especially at the disk periphery, compared to configurations where $i = 10\degree$.

\begin{figure}[!ht]
\centering\includegraphics[width=0.6\textwidth]{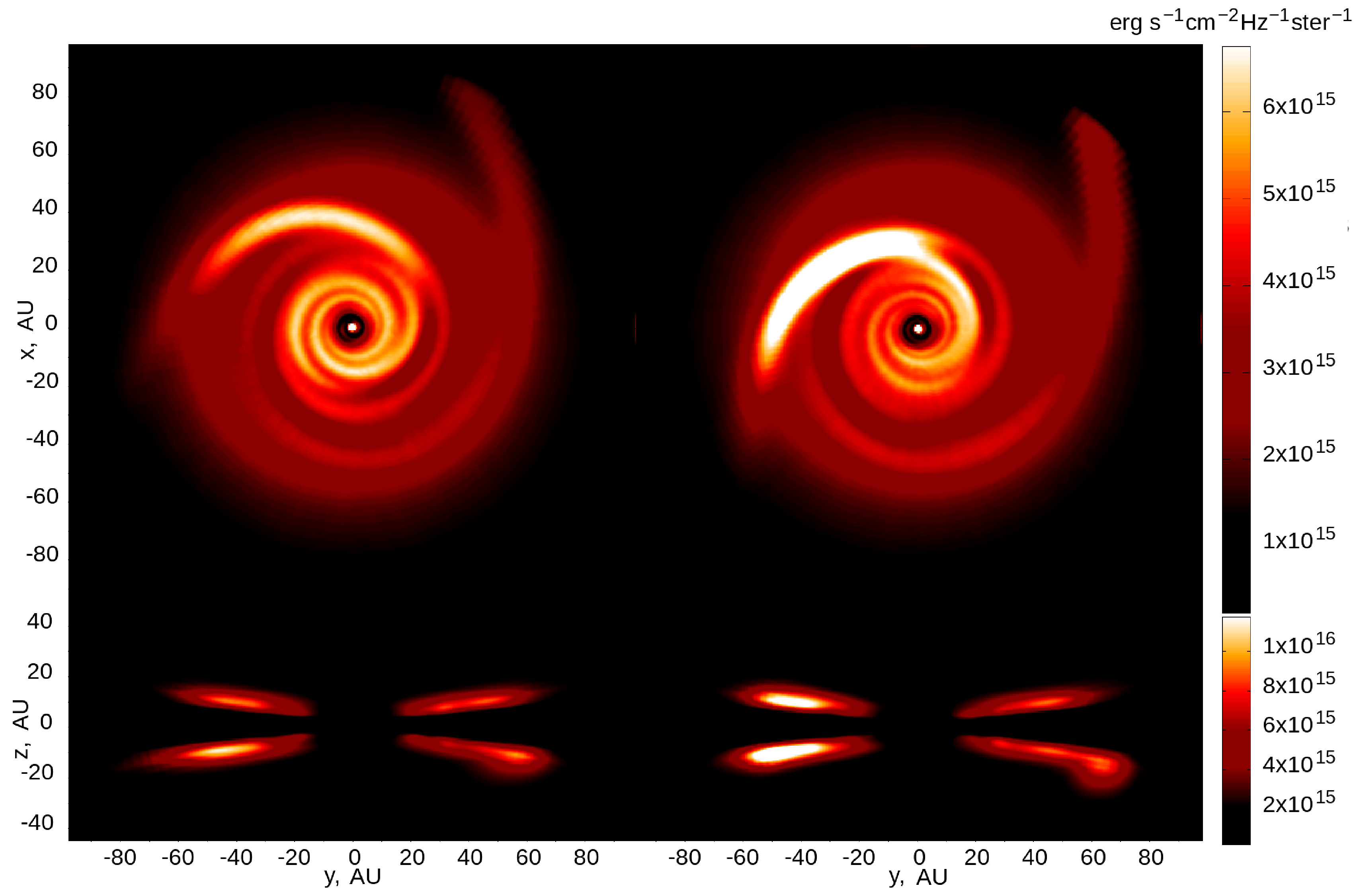}
\caption{\rm Same as in Fig.~\ref{fig:timeP} for the case of a prograde flyby of the planet at time $t=290$~years.
On the left is a model in which the pericentric distance is $q=10$~AU, and on the right is $q=20$~AU ($i=10\degree$, $\omega=90\degree$). In the lower graphs, the line of sight is directed along the $x$ axis.}\label{fig:qP}
\end{figure}

\begin{figure}[!ht]
\centering\includegraphics[width=0.6\textwidth]{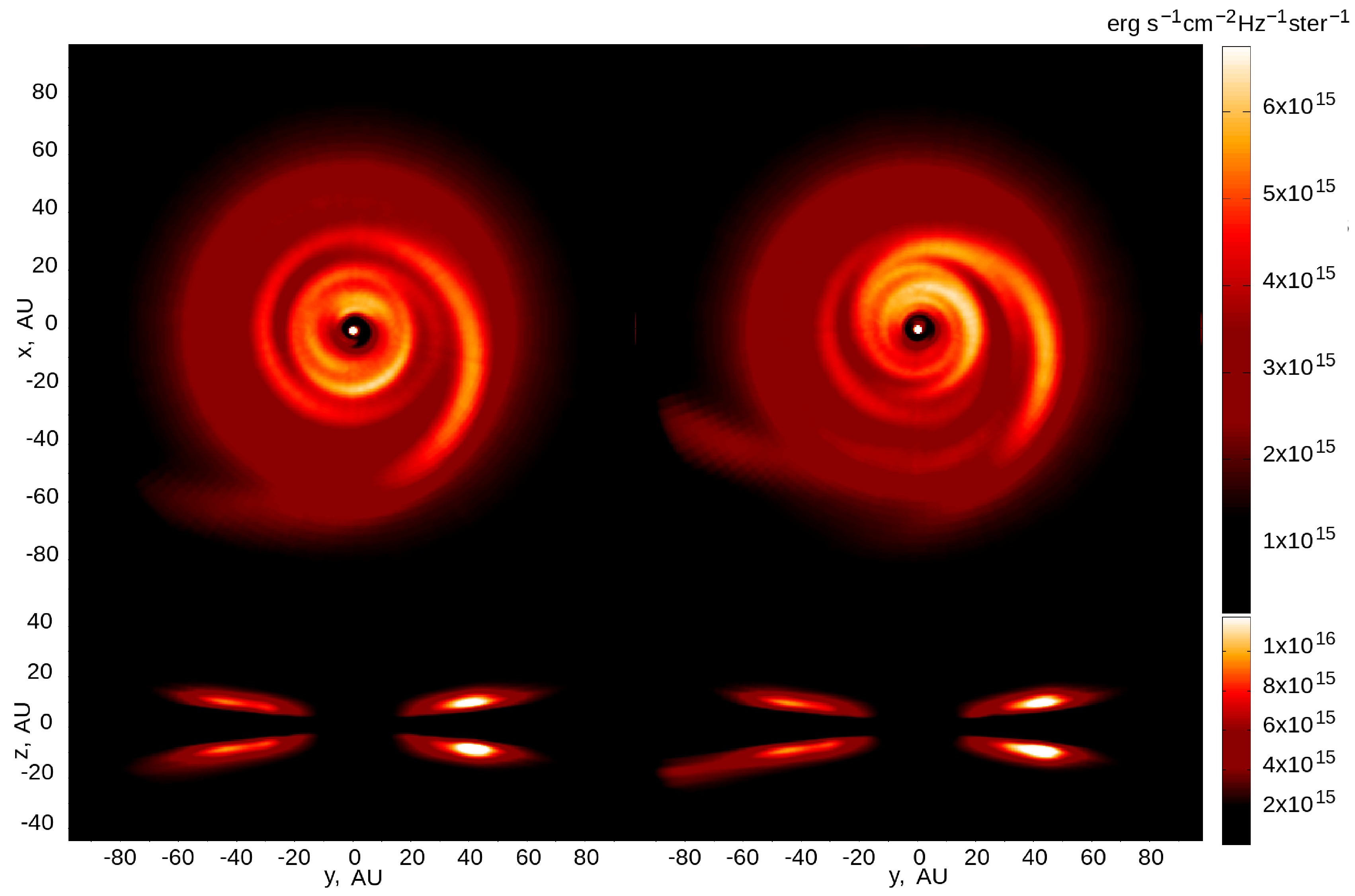}
\caption{\rm The same as in Fig.~\ref{fig:qP} for the case of a retrograde flyby of the planet.}\label{fig:qR}
\end{figure}

As the pericentric distance $q$ increases (within the disk size), the probability of a collision between the disk and the planet increases. Calculations have shown, as expected, that the farther from the star the interaction occurs, the more the rarefied periphery of the disk is distorted (Fig.~\ref{fig:qP}, \ref{fig:qR}). Spiral structures appear brighter, and tails of matter extend from the disk in the direction of the planet's motion. 

In the model considered above, the first and second intersections of the disk plane by the planet occur at the same distance ($r\approx 10$~AU), since $\omega = 90\degree$. Three more cases were considered: first, a single intersection of the disk plane by the planet's orbit ($\omega=0\degree$), and second, when the first intersection ($r_1$) occurs closer to the star than the second ($r_2$) and vice versa. When $\omega=45\degree$: $r_1\approx 5.87$~AU, $r_2\approx 34.14$~AU; and in the case of $\omega=135\degree$: $r_1\approx 34.14$~AU, and $r_2\approx 5.87$~AU. In general, the picture of inhomogeneities differs little with variations in $\omega$. However, it can be noted that in the case of $\omega=0\degree$ and $\omega=135\degree$ the periphery of the disk is more disturbed, whereas at $\omega=45\degree$ one branch of the spiral is significantly brighter than the other during the prograde flyby. And in the retrograde case the single-arm spiral is brighter at $\omega=0\degree$ and $\omega=45\degree$.

Reducing the planet's mass by an order of magnitude ($1M_{\rm Jup}$) reduces the height to which the disturbed disk matter rises. Although spiral structures similar to those described above stand out in the gas distribution, they are indistinguishable from the rest of the disk when modeling images. The minimum mass at which spiral structures can be identified is $3M_{\rm Jup}$.

\begin{figure}[!ht]
\centering\includegraphics[width=0.6\textwidth]{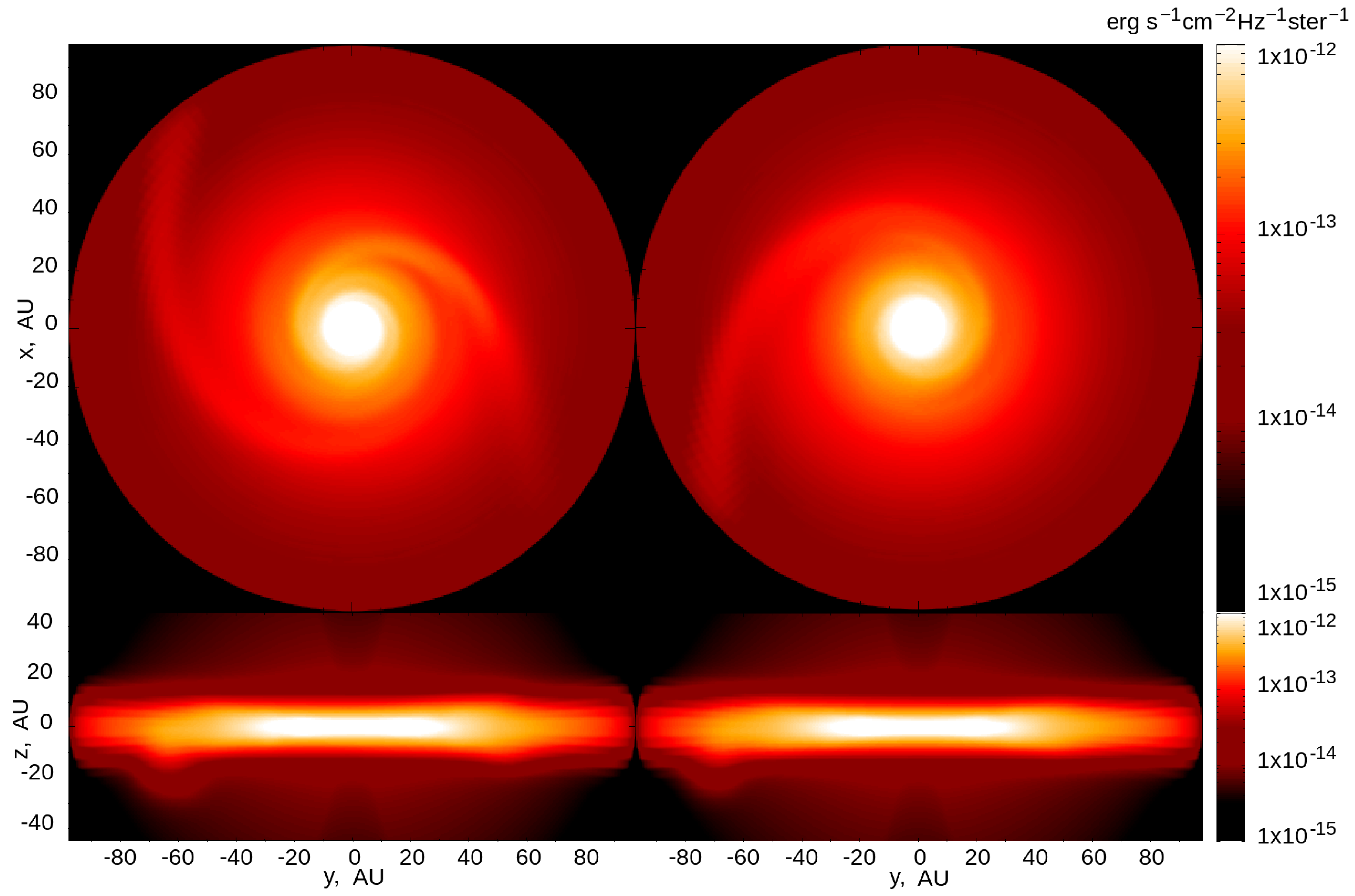}
\caption{\rm The color shows the radiation flux at a wavelength of $740$~$\mu$m in a logarithmic scale at time $t=290$~years.
On the left is a prograde flyby for the model with parameters: $q=20$~AU, $i=10\degree$, $\omega=90\degree$, and on the right is a retrograde one for the model: $q=5$~AU, $i=10\degree$ and $\omega=0\degree$. The upper graphs correspond to the direction of the line of sight against the $z$ axis, and the lower ones along the $x$ axis.}\label{fig:l740}
\end{figure}

\begin{figure}[!ht]
\centering\includegraphics[width=1.0\textwidth]{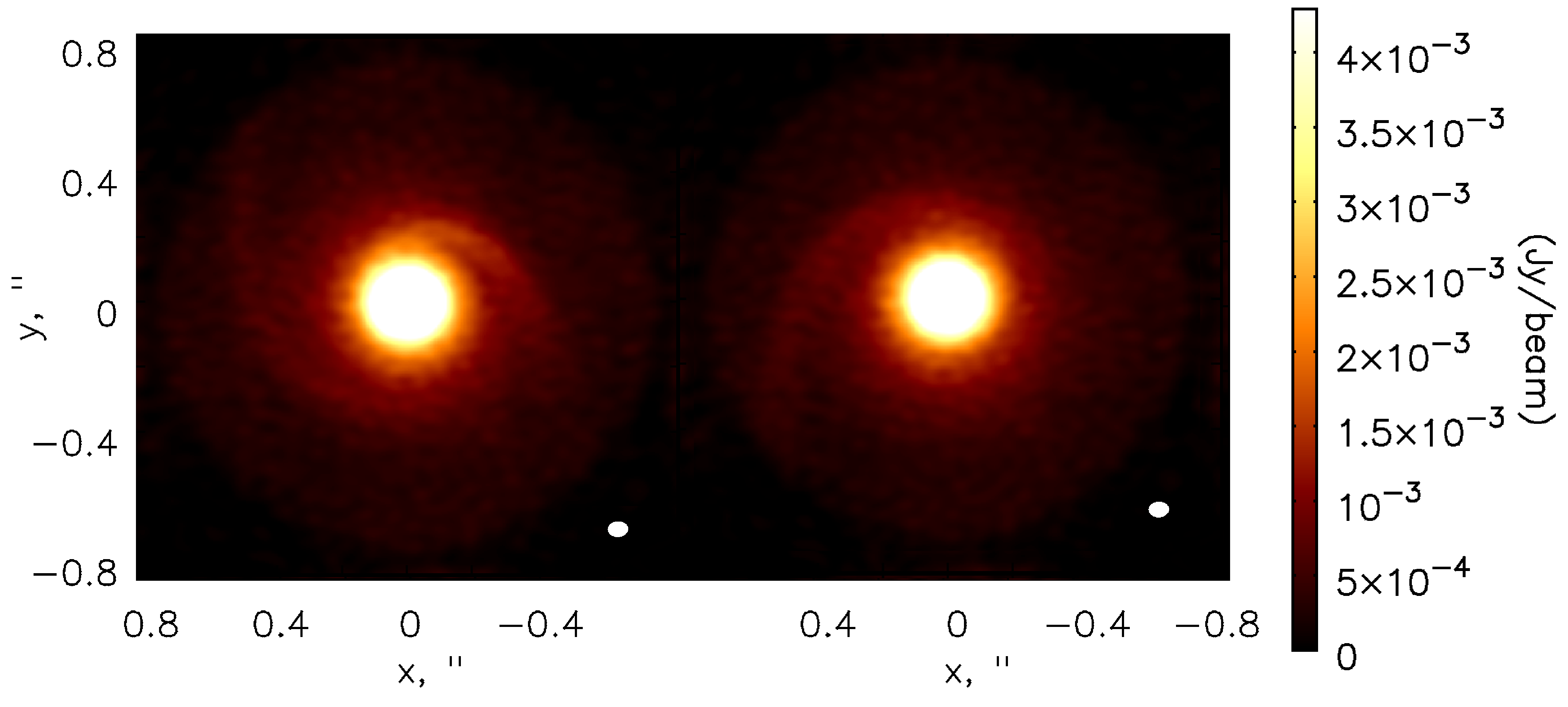}
\caption{\rm The synthesized images that could potentially be obtained with the ALMA radio interferometer at a wavelength of 740 $\mu$m correspond to the models presented in Fig.~\ref{fig:l740}. The color scale is given in Jy/beam.}\label{fig:alma}
\end{figure}

Analysis of the thermal emission of dust showed that the above-described structures can be observed if the mass of fine dust
(0.1 $\mu$m in size) is at least $10^{-4} M_\odot$. Fig.~\ref{fig:l740} shows the radiation flux for two models with the brightest two-arm and one-arm spirals relative to the background. On a logarithmic scale, the spiral structures stand out against the background of the rest of the disk, and the distortion of the disk plane from the edge is also visible. The spiral structures are also distinguishable in theoretical images obtained for the ALMA telescope (Fig.~\ref{fig:alma}), while the vertical distortions of the disk are not identified. Calculations have shown that observations of single-arm spirals in the submillimeter range are possible in the case of a close flyby at $q\approx 5$~AU, while double-arm spirals can be resolved in a non-close flyby with a pericentric distance of $q\approx 20$~AU.

\section*{DISCUSSION}
\noindent
The influence of a flyby of a solar-type star on the dynamics of a protoplanetary disk has been considered in a number of studies. Calculations by~\citet{2008A&A...492..735P} showed that such an event can cause a burst of accretion activity on the star, during which the accretion rate increases by several orders of magnitude. The prograde flyby of the star through disk matter significantly changed the structure of the disk, a tidal spiral tail of matter was formed, and part of the matter was captured by the incoming star~\citep{2010MNRAS.402.1349F}.

Calculations taking into account the interaction of gas and dust~\citep{2019MNRAS.483.4114C} showed that the close flyby of the star provokes the development of a spiral structure in the distribution of dust and gas, which persists for thousands of years after periastron. The plane of the disk is distorted, and the dust disk becomes more compact compared to the gas disk. Images of these disks were analyzed in the optical, infrared and submillimeter ranges~\citep{2020MNRAS.491..504C}.
	
In the paper by~\citet{2022MNRAS.510L..37B}, similar calculations were repeated taking into account the radiative transfer. They showed that the accretion rate onto the incident star can reproduce a FU~Ori-like outburst in terms of timing and amplitude. However, the star should fly through the central regions of the disk ($<20$~AU). Such encounters are unlikely even in star clusters. A more distant flyby of a star with a pericentric distance of $500$~AU can provoke the development of thermal and magneto-rotational instability in the inner part of the disk, which can lead to a delayed ($\sim 1000$~years after the encounter) burst of the accretion rate~\citep{2023ARep...67.1401S}.
	
The case of a penetrating encounter of two stars with protoplanetary disks was considered in~\citet{2022MNRAS.517.4436B}. Calculations showed that spiral waves propagate in the disks of both stars, the plane of both disks is distorted, the sizes of the disks decrease after the flyby, they are connected by a spiral tail of matter. Thus, in the case of an approach of a star to a protoplanetary disk, its structure changes~\citep{2023EPJP..138...11C}. Spiral waves are formed, the plane of the disk is deformed, its eccentricity increases, and some of the matter can be captured by the incoming star. A prograde flyby has a more destructive effect on the protoplanetary disk than a retrograde one.
	
During a retrograde approach to a massive star ($\geq 5 M_\odot$), the protoplanetary disk can change the tilt of its plane relative to the equatorial plane of the parent star~\citep{2016MNRAS.455.3086X}.

In~\citet{2020A&A...635A.196V}, the processes of pulling out tails of matter (tail-like structures) from protoplanetary disks were studied. The authors showed that the ejection of clumps from massive gas disks of young stars due to gravitational instability can also cause the formation of tail-like structures, which are shock waves with a dense perimeter and a rarefied interior. In addition, when a star flies close through a gravitationally stable disk, three tails of matter can form. Before and after the collision with the disk, the incoming star drags matter from the disk, and a spiral massive tail of matter is formed. Such structures can potentially be observed in scattered light and the submillimeter range. Extended flows of matter associated with protoplanetary disks have been detected in a number of young objects~\citep{2006A&A...452..897C,2015ApJ...806L..10D,2022A&A...658A.104G}.

In the previous paper, where we modeled the interaction of a gas stream with a protoplanetary disk~\citep{2024ARep...68..949G}, the disk distortions were also significant already at the initial mass of the stream equal to 1 Jupiter mass. Spiral waves were observed, an inner disk was formed, tilted relative to the periphery, which preserved the initial plane of the disk.

In this paper, we have shown that when a massive planet flies through a protoplanetary disk, the structure of the protoplanetary disk changes less significantly. However, similar structures, as in the case of another star flying through, are also observed when the disk approaches the planet. First, a single-arm or double-arm spiral is visible in the disk images from the pole. Second, the plane of the disk is distorted, and finally, a tail of matter extends from the disk in the direction of the planet. It should be noted that in the models considered, due to the limited size of the computational domain, this tail is cut off, and in a number of models, its existence is expected beyond the computational grid.

Spirals were detected in images of the stars AB Aur~\citep{2004ApJ...605L..53F}, Elias 2-27~\citep{2021ApJ...914...88P}, MWC 758~\citep{2018ApJ...860..124D}, HD 135344B~\citep{2021MNRAS.507.3789C}. Spiral waves identified in images of protoplanetary disks can be caused by various reasons. For example, the motion of the planet along a closed orbit or the development of gravitational instability~\citep[see, for example,][]{2017ApJ...839L..24M}. However, the flyby of a planet through a protoplanetary disk can also give rise to such structures. The formation of transient spiral structures was also revealed in the modeling of circumbinary planetesimal disks~\citep{2015ApJ...805...38D}.

A number of seen edge-on objects also show a noticeable disk asymmetry relative to the central plane. This is probably due to a distortion of the disk plane. Examples are the stars HD 111520~\citep{2022ApJ...932...23C}, Proplyd 114-426~\citep{1996AJ....111.1977M}, IRAS 23077+6707~\citep{2024ApJ...967L...2M}, Oph 163131~\citep{2020A&A...642A.164V,2021AJ....161..239F}, HD 141943~\citep{2014ApJ...786L..23S}. It is interesting to note the star HD106906, whose protoplanetary disk is observed edge-on. On one side, the edge of the disk is more elongated than on the other, and resembles a tail of matter that stretches behind the planet. In addition, a distant planet has also been discovered, likely in a highly elliptical orbit~\citep{2021AJ....161...22N}. 

\section*{CONCLUSION}
Calculations have shown that a massive planet of 10 Jovian masses can significantly disturb the matter of the protoplanetary disk.
When such a planet flies through the disk in a prograde direction, two spiral humps appear on its surface,
and when it flies in a retrograde direction, only one. These humps are visible in disk images as spiral structures when observed pole-on, and as asymmetries relative to the plane of the disk when observed edge-on. They can be identified both in the IR and in the submillimeter range. In addition, the planet can stretch a tail of matter from the disk, which can also be detected during observations. Observation of such structures is possible in nearby star-forming regions at distances of $\sim 140$~pc.

Variations in the planet's orbital parameters affect the brightness of individual structures, but the overall pattern of large-scale inhomogeneities remains. Over time, the structures dissipate in the disk with a characteristic time of $\sim 500$~yr. This time interval is significantly shorter than the characteristic lifetime of a protoplanetary disk of $\sim 10^6$~yr~\citep{2001ApJ...553L.153H,2011ARA&A..49...67W}, so the consequences of a single flyby are unlikely to be detected during observations. However, the results of the studies are also applicable to the case of a planet on a highly elongated elliptical orbit. For example, the orbit of a planet with a pericentric distance of $5$~AU and an apocentric distance of $150$~AU has a semi-major axis of $77.5$~AU, eccentricity of $\approx 0.94$~AU. In this case, the period of revolution around a Sun-like star is $\approx 682$~years, therefore, during the lifetime of the protoplanetary disk, such a planet will collide with the disk several thousand times. This significantly increases the probability of observing the described structures.

Analysis of calculations showed that the flyby of the planet does not affect the rate of gas accretion from the disk onto the star. Such a phenomenon cannot lead to the formation of an inclination of the inner part of the disk relative to the periphery. However, under certain conditions, it can cause an eclipse on the light curve of a UX Ori-type star~\citep{2025LJM...46..95G}. 

The results obtained are also applicable in the case of a debris disk, since a massive planet, first of all, exerts a gravitational effect on the disk. In addition, if the planet moves along a highly elongated orbit that intersects the plane of the disk, the structures described above should arise in it and then dissipate. Thus, the probability of observing the described structures will be higher, since they can arise periodically, albeit with a fairly large period.
 
\section*{Acknowledgments}
The authors are grateful to the reviewers for their valuable comments.
The calculations were carried out using the resources of Joint Supercomputer Center (JSCC).

\bibliographystyle{rusnat}
\bibliography{GD24}

\end{document}